\documentclass[12pt]{article}
\usepackage{amsmath,amssymb}
\usepackage{epsfig} 

\def\m@th{\mathsurround=0pt }
\def\leftrightarrowfill{$\m@th \mathord\leftarrow \mkern-6mu
        \cleaders\hbox{$\mkern-2mu \mathord- \mkern-2mu$}\hfill
        \mkern-6mu \mathord\rightarrow$}

\def\overleftrightarrow#1{\vbox{\ialign{##\crcr
        \leftrightarrowfill\crcr\noalign{\kern-1pt\nointerlineskip}
        $\hfil\displaystyle{#1}\hfil$\crcr}}}

\def\simlt{\stackrel{<}{{}_\sim}}
\def\simgt{\stackrel{>}{{}_\sim}}

\newcommand{\be}{\begin{equation}}
\newcommand{\ee}{\end{equation}}

\def\shat{\ifmmode \hat{s}\else $\hat{s}$\fi}
\def\gp2{{g'}^2}
\def\g2{g^2}
\def\g32{g_s^2}

\newcommand{\newc}{\newcommand}

\newc{\gsim}{\lower.7ex\hbox{$\;\stackrel{\textstyle>}{\sim}\;$}}
\newc{\lsim}{\lower.7ex\hbox{$\;\stackrel{\textstyle<}{\sim}\;$}}
\newc{\ie}{{\it i.e.}}
\newc{\etal}{{\it et al.}}
\newc{\mev}{\hbox{\rm\,MeV}}
\newc{\gev}{\hbox{\rm\,GeV}}
\newc{\tev}{\hbox{\rm\,TeV}}
\newc{\xpb}{\hbox{\rm\, pb}}
\newc{\xfb}{\hbox{\rm\, fb}}

\newc{\G}{{\cal G}}
\newc{\h}{{\cal H}}
\newc{\D}{{\cal D}}
\newc{\E}{{\cal E}}

\newc{\mtop}{M_t}
\newc{\mbot}{m_b}
\newc{\mz}{M_Z}
\newc{\mw}{M_W}
\newc{\alphasmz}{\alpha_s(M_Z)}
\newc{\swsq}{\sin^2\theta_W}
\newc{\cwsq}{\cos^2\theta_W}
\newc{\tw}{\tan\theta_W}
\newc{\cw}{\cos\theta_W}
\newc{\sw}{\sin\theta_W}
\newc{\BR}{\hbox{\rm BR}}
\newc{\zbb}{Z\to b\bar}
\newc{\Gb}{\Gamma (Z\to b\bar b)}
\newc{\Gh}{\Gamma (Z\to \hbox{\rm hadrons})}
\newc{\sgn}{\mbox{sgn}}

\def\eq#1{eq.~(\ref{#1})}

\newcounter{mysubequation}[equation]

\def\beq{\begin{equation}}
\def\eeq{\end{equation}}
\def\bea{\begin{eqnarray}}
\def\eea{\end{eqnarray}}
\def\slashchar#1{\setbox0=\hbox{$#1$}           
   \dimen0=\wd0                                 
   \setbox1=\hbox{/} \dimen1=\wd1               
   \ifdim\dimen0>\dimen1                        
      \rlap{\hbox to \dimen0{\hfil/\hfil}}      
      #1                                        
   \else                                        
      \rlap{\hbox to \dimen1{\hfil$#1$\hfil}}   
      /                                         
   \fi}                                         %
\catcode`@=11
\long\def\@caption#1[#2]#3{\par\addcontentsline{\csname
  ext@#1\endcsname}{#1}{\protect\numberline{\csname
  the#1\endcsname}{\ignorespaces #2}}\begingroup
    \small
    \@parboxrestore
    \@makecaption{\csname fnum@#1\endcsname}{\ignorespaces #3}\par
  \endgroup}
\catcode`@=12

\begin{document}

\baselineskip=18pt

\setcounter{footnote}{0}
\setcounter{figure}{0}
\setcounter{table}{0}

\begin{titlepage}
\begin{flushright}
CERN-PH-TH/2008--059
\end{flushright}
\vspace{.3in}

\begin{center}
{\Large \bf Supersymmetric Leptogenesis\\
 and
the Gravitino Bound}

\vspace{0.5cm}

{\bf G.F. Giudice$^{a}$}, {\bf L. Mether$^{a,b}$},  
{\bf A. Riotto$^{a,c}$} and {\bf F. Riva$^{a,d}$}
\vskip 0.5cm

\centerline{$^{a}${\it CERN, Theory Division, CH--1211 Geneva 23, Switzerland}}
\centerline{$^{b}${\it Helsinki Institute of Physics, 
P.O. Box 64, FIN-00014, Helsinki, Finland}}
\centerline{$^{c}${\it INFN, Sezione di Padova, Via Marzolo 
8, I-35131 Padua, Italy}}
\centerline{$^{d}${\it Rudolfs Peierls Centre for 
Theoretical Physics, University of Oxford,} }
\centerline{{\it 1 Keble Rd., Oxford OX1 3NP, UK}}

\end{center}
\vspace{.8cm}

\begin{abstract}
\medskip
\noindent
Supersymmetric thermal leptogenesis with a hierarchical right-handed
neutrino mass spectrum requires the mass of the lightest right-handed
neutrino to be heavier than about $10^9$ GeV. This is in conflict with the
upper bound on the reheating temperature 
which is found by imposing that the gravitinos generated
during the reheating stage after inflation do not jeopardize successful
nucleosynthesis. In this paper we show that a solution to this tension
is actually already incorporated in the framework, because of the presence
of flat directions in the supersymmetric scalar potential.
Massive 
right-handed neutrinos
are efficiently produced non-thermally and the observed baryon asymmetry
can be explained even for a reheating temperature respecting the gravitino
bound if  two conditions are satisfied: 
the initial value of the flat direction must be 
close to Planckian values and the
phase-dependent terms in the flat direction potential are either vanishing
or sufficiently small.

\end{abstract}

\bigskip
\bigskip

\end{titlepage}

\noindent
The observed baryon number asymmetry (normalized with respect to the 
 entropy density)
of the Universe $Y_B=\left(0.87\pm 0.03\right)
\times 10^{-10}$ \cite{wmap} can be explained by 
the mechanism of thermal leptogenesis \cite{FY,review}, 
the simplest implementation of this mechanism being realised by adding to 
the Standard Model (SM)
three heavy right-handed (RH) neutrinos. In thermal leptogenesis  
the heavy RH neutrinos
are produced by thermal scatterings after inflation  and subsequently
decay out-of-equilibrium in a lepton number and CP-violating way.
The 
dynamically generated  lepton asymmetry is 
then  converted into a baryon asymmetry
due to $(B+L)$-violating sphaleron interactions \cite{kuzmin}.

If  RH neutrinos are hierarchical in mass,
successful leptogenesis requires that the mass $M_1$ of
the lightest RH neutrino $N_1$ 
is  larger than $2\times 10^9$ GeV, for vanishing initial $N_1$ 
density~\cite{di}. This lower limit on $M_1$ is reduced to 
$5\times 10^8$~GeV when $N_1$ is initially in thermal equilibrium 
and to $2\times 10^7$~GeV when $N_1$ initially dominates the energy 
density of the Universe~\cite{strum}.
 These results do not substantially change when flavour 
effects are accounted for
\cite{f1}. 
Hence, in the standard framework of thermal leptogenesis, 
the required reheating temperature after inflation $T_{\rm RH}$ 
cannot be lower than about $2\times 10^9$~GeV~\cite{strum}. In 
supersymmetric scenarios this is in conflict with the upper bound
on the reheating temperature necessary to 
avoid the  overproduction of gravitinos during
reheating \cite{grav}. Being only gravitationally coupled to 
SM particles (and their supersymmetric partners), gravitinos decay
very late jeopardizing the successful predictions  of Big Bang nucleosynthesis.
This does not happen, however, if gravitinos are not efficiently
produced during reheating, that is if the reheating temperature 
$T_{\rm RH}$ is small enough. For gravitino masses in the natural range 
from 100 GeV to 1~TeV, within the minimal supergravity framework, 
the reheating tempeature  should be smaller than about $10^5$--$10^7$ 
GeV~\cite{grav}, depending on the chosen values of the supersymmetric parameters and of the primordial element abundances.

The severe bound on the reheating temperature makes the thermal generation
of the RH neutrinos impossible, thus rendering the supersymmetric 
thermal leptogenesis scenario not viable if RH neutrinos are hierarchical. 
Of course, there are  several ways out to this drawback. 
First of all, one can modify the usual assumptions on gravitinos. If the gravitino is stable, the nucleosynthesis limit depends on the nature of the next-to-lightest supersymmetric particle, but values of $T_{\rm RH}$ even larger than $10^9$ GeV can be obtained~\cite{stable}. Assuming the existence of small $R$-parity violation, the next-to-lightest supersymmetric particle can decay before the onset of supersymmetry, evading the bound on $T_{\rm RH}$~\cite{ibarra}. Also, gravitinos lighter than 1 KeV (as possible in gauge mediation) or heavier than about 50 TeV (as possible in anomaly mediation) avoid the stringent limits on $T_{\rm RH}$. 
Alternatively, one can modify the standard mechanism of leptogenesis, and rely on 
supersymmetric resonant leptogenesis~\cite{res} or soft leptogenesis~\cite{soft}.
Indeed, in resonant leptogenesis the RH neutrinos are nearly degenerate in mass and self-energy 
contributions to the CP asymmetries are enhanced, thus producing the correct baryon asymmetry even at  temperatures as low as the
TeV.  Soft leptogenesis  can be successful for values of the mass $M_1$ of 
the lightest RH neutrino as low as $10^6$ GeV. Another interesting variation is the case in which the right-handed sneutrino develops a large amplitude, dominating the total energy density~\cite{murayy}. Then the sneutrino decay reheats the universe, producing a lepton asymmetry, where values of $T_{RH}$ as low as $10^6$~GeV do not cause a gravitino problem.
Finally, one can modify the standard thermal production mechanism of $N_1$.
The lightest RH neutrinos can be produced
 non-thermally either during the preheating stage
\cite{noi}, or from the inflaton decays \cite{decay} 
or from quantum fluctuations
\cite{noi2}. 

In this paper, we would like to show that a solution to the tension 
between supersymmetric
leptogenesis with hierarchical RH neutrinos 
and the gravitino bound is in fact already rooted
in one of the basic properties of the supersymmetric theory, that is 
the  presence of flat directions in the scalar potential \cite{karirev}. 
No new ingredient has to be added to the theory.
Let us briefly 
sketch how the solution works. The $F$- and $D$-term 
flat directions are lifted because of the presence of  the soft
supersymmetry breaking terms in our vacuum, of possible
non-renormalizable terms in the superpotential and 
of  finite energy density  terms in the potential
proportional to the Hubble rate $H$ during inflation~\cite{dinesusy}. As
a consequence, the field $\phi$  along the flat direction
will acquire a large vacuum expectation value (VEV). 
When, after inflation, the Hubble rate becomes of the order of the supersymmetry
breaking mass $\widetilde{m}$, the condensate starts oscillating around
the true minimum of the potential which resides at $\phi=0$. 
If the condensate passes close enough to the origin, the  particles
coupled to the condensate are efficiently created at the first 
passage. The produced particles become
massive once the condensate continues its oscillation leaving 
the origin and may efficiently decay into other massive states, in our case
RH neutrinos. The latter will subsequently decay to generate the final
baryon asymmetry.
The  process allowing the generation of very massive states is called  instant preheating 
\cite{lindeinstant} and represents a very efficient way of producing 
heavy states. In this sense, the solution we are proposing 
may be considered as a non-thermal production of RH neutrinos, but 
we stress that it does not involve any extra assumption such as 
a large coupling between the RH neutrinos and the inflaton field.

The generic potential for a supersymmetric flat direction $\phi$ is given
by \cite{dinesusy}
\begin{equation}
\label{pot}
V(\phi)=\left(\widetilde{m}^2-c H^2\right)\left|\phi\right|^2
+\left(\lambda\frac{A+a H}{n M^{n-3}}\phi^n
+{
\rm h.c.}\right)+
\left|\lambda\right|^2\frac{\left|\phi\right|^{2n-2}}{M^{2n-6}},
\end{equation}
where $c$, $a$ and $\lambda$ 
are constants of ${\cal O}(1)$, $\widetilde{m}$ and $A$ are
the soft breaking mass terms of order the TeV scale, 
$H$ is the Hubble rate, $M$ is some large 
mass scale which we assume to be equal to the reduced the Planck scale 
($M=M_p=2.4\times 10^{18}$~GeV) and $n$ is an integer 
larger than three. 
For $c>0$ and $H\gg \widetilde{m}$,  the flat direction
condensate acquires a VEV given by 
\begin{equation}
\left|\phi_0\right|=\left(\frac{\beta H M^{n-3}}{\lambda}\right)^{1/(n-2)},
\label{phi0}
\end{equation}
where $\beta$ is a numerical constant which depends on $a$, $c$, and $n$.  
At the end of inflation, the inflaton field starts oscillating around the 
bottom of its potential and the Hubble rate  decreases. As soon as  
$H\sim\widetilde{m}/3$, the condensate starts rolling down towards its minimum
at $\phi=0$. 

Now, if in the potential in eq.~(\ref{pot}) both terms proportional
to $A$ and $a H$ are present and their relative phase $\theta_a-\theta_A$ 
does not vanish, 
the condensate $\left|\phi\right|e^{i\theta}$ will spiral around  
the origin at $\phi=0$ with a nonvanishing
$\dot{\theta}$ (possibly leading to a large baryon asymmetry through the
Affleck-Dine mechanism \cite{ad,dinesusy}). In this case instant preheating does not occur and no heavy states are produced~\cite{noinst}, unless several flat directions are simultaneously excited~\cite{peloso}. We will focus on the opposite case, when the condensate passes
through the origin (or sufficiently close to it). 
This is easy  to achieve without  any fine-tuning \cite{dinesusy}
as it is  enough
to consider a flat direction which is lifted by a non-renormalizable
superpotential term which contains a single field not in the flat direction
and some number of fields which make up the flat direction \cite{dinesusy},
\begin{equation}
W=\frac{\lambda}{M^{n-3}}\psi\phi^{n-1}.
\end{equation}
For terms of this form, $F_\psi$ is non-zero along the flat direction, but
$W=0$ along it. Examples of this type are represented by the direction
$ue$ which is lifted by $W=(\lambda/M)uude$, since $F^*_d=(\lambda/M)uue$ 
is non-zero along the direction, and by the $Que$ direction which is lifted
by the $n=9$ superpontial $W=(\lambda/M)QuQuQuH_Dee$ since  
$F^*_{H_D}=(\lambda/M)QuQuQuH_Dee$ does not vanish \cite{gh}.  
If $W=0$ along the flat direction, no phase-dependent
terms are induced.   
Alternatively, the superpotential may vanish 
along the flat direction because of
a discrete $R$-symmetry.  In such a case, when $W$ exactly vanishes, 
the potential
during inflation has the form \cite{dinesusy}
\begin{equation}
V(\phi)=H^2 M_p^2 f(\left|\phi\right|^2/M_p^2)+H^2 M_p^2 g(\phi^n/M_p^n),
\end{equation}
and the typical initial value $\phi_0$ 
for the condensate is  ${\cal O}(M_p)$, rather than \eq{phi0}. For this reason we will treat $\phi_0$ essentially as a free parameter 
in our analysis and not fixed by the relation \eq{phi0}. Finally we remark that the coefficients $A$ and $a$ depend on the specific form of the K\"ahler potential couplings and there are cases in which they are suppressed by inverse powers of $M_p$. For instance, if the inflaton is a composite field, it will appear in the K\"ahler potential only through bilinear combinations and $a\sim H/M_p$. In the case of $D$-term inflation \cite{dterm} 
$a$ vanishes identically
and no phase-dependent terms are generated if along the flat direction
$W=0$.

From now on, we will consider a flat direction along which the induced $A$ terms are suppressed and therefore the corresponding condensate will oscillate passing very close to the origin. Furthermore, we 
will focus  on  the flat direction involving the third generation quark $u_3$. 
When the condensate
passes through the origin, it can efficiently produce states which are coupled
to it. Let us consider the scalar Higgs $H_U$ which is relevant for 
leptogenesis although, of course, other states will be produced as well. 
If the third generation is
involved in the flat direction, 
the up-Higgs  is coupled to
the condensate through the Lagrangian term $h_t^2\left|\phi\right|^2
\left|H_U\right|^2$. Its effective mass is therefore given by 
$m_{H_U}^2=\widetilde{m}^2_{H_U}+h_t^2 \left|\phi\right|^2$, where 
$\widetilde{m}^2_{H_U}$ is the corresponding soft-breaking mass parameter. At the
first passage through the origin, particle production takes place when 
adiabaticity is violated \cite{lindeinstant}, 
$\dot{m}_{H_U}/m^2_{H_U}\simgt 1$. This requires 
\begin{equation}
\frac{|\dot{\phi}|}{h_t\left|\phi\right|^2}\sim
\frac{\widetilde{m}|\phi_0|}{h_t\left|\phi\right|^2}\simgt 1.
\end{equation}
Up-Higgses can therefore be efficiently produced if $\left|\phi\right|\simlt
\left(\widetilde{m}|\phi_0|/h_t\right)^{1/2}\equiv \left|\phi_*\right|$. 
As a result, particle production occurs nearly instantaneously, within
a time
\begin{equation}
\Delta t_*\sim \frac{\left|\phi_*\right|}{|\dot{\phi}|}\sim \left(h_t
\widetilde{m}|\phi_0|\right)^{-1/2}.
\end{equation}
The uncertainty principle implies that the created up-Higgses are generated
with typical momentum \cite{lindeinstant}
\begin{equation}
k_*\sim \left(h_t
\widetilde{m}|\phi_0|\right)^{1/2}
\end{equation}
and with a number density 
\begin{equation}
n_{H_U}\sim\frac{k_*^3}{8\pi^3}\sim \frac{\left(h_t
\widetilde{m}|\phi_0|\right)^{3/2}}{8\pi^3}.
\label{nhu}
\end{equation}
After the condensate has passed through the origin  
 continuing its motion,
the up-Higgses become heavier and heavier, having an effective  mass 
$\sim h_t\left|\phi\right|$. When this mass becomes
larger than the  lightest RH neutrino mass $M_1$, the up-Higgses
will promptly decay into the RH neutrinos $N_1$ 
(we suppose that the 
other RH neutrinos are much heavier than $M_1$) 
through the superpotential 
coupling $h_{ij} N_i \ell_j H_U$, where $\ell_j$ 
stands for the lepton doublet of
flavour $j$ and $i,j=1,2,3$.
Indeed, the $H_U$ decay is prompt because the decay rate 
$\Gamma_D \sim \sum_j |h_{1j}|^2 h_t \phi /(8\pi)$ is faster than 
the oscillation rate $\Gamma_{\rm osc} \sim \dot \phi /\phi$ 
as long as $\phi^2>8\pi \widetilde{m} \phi_0 /(\sum_j |h_{1j}|^2 h_t )$, 
which is certainly satisfied during the first oscillation. 
Moreover, if one of the  $h_{1j}$ is   not too small, and $Q_3$ is not involved 
in the flat direction\footnote{For the $Que$ flat direction the $n=9$ lifting 
superpotential contains $Q_3$ only if all the   $n=4$ lifting superpotentials
$QQQL$, $QuQd$, $QuLe$ and $uude$
are  present in the supersymmetric Lagrangian.}, $H_U$ will dominantly decay into $N_1 \ell$, 
since any decay process occurring through top-Yukawa or gauge interaction 
is kinematically forbidden (or strongly suppressed) at large $\phi$. 

To estimate the maximum value $M_1^{\rm max}$ that can be generated we have to compute the maximum value $\phi^{\rm max}$ achieved by the condensate during its first oscillation, after passing through the origin. The equation of motion for $\phi$ is
\begin{equation}
\label{9}
\ddot{\phi}+\widetilde{m}^2\phi=-h_t \frac{|\phi|}{\phi}n_{H_U}.
\end{equation}
The term on the right-hand side corresponds to the $\phi$-dependent energy density $m_{H_U}(\phi)n_{H_U}$ generated by the $H_U$ particles produced when $\phi$ crosses the origin. It acts as a friction term damping the $\phi$ oscillations. Solving \eq{9}, we obtain
\beq
M_1^{\rm max}\simeq h_t \phi^{\rm max} = \frac{4\pi^3 \widetilde{m}^{1/2} \phi_0^{1/2}}{h_t^{3/2}} =
4\times 10^{12}~{\rm GeV} \left(\frac{\phi_0}{M_p}\right)^{1/2}\left( \frac{\widetilde{m}}{100 \,{\rm GeV}}\right)^{1/2},
\label{mmax}
\eeq
where we have taken the top-Yukawa coupling $h_t\simeq 0.6$ at high-energy scales. Thus, very heavy RH neutrinos can be produced through this mechanism.

  In first approximation, we can assume that all $H_U$ decay into $N_1$ and the number density of the
RH neutrinos is given by $n_{N_1}\sim n_{H_U}\sim \left(h_t
\widetilde{m}|\phi_0|\right)^{3/2}/8\pi^3$. When the mass of the up-Higgses 
decreases because the condensate, after reaching its maximum value at the first
oscillation, starts decreasing again, the RH neutrinos may efficiently
decay into up-Higgses and leptons and produce a lepton asymmetry $n_L\sim
\epsilon n_{N_1}$ where
the usual CP asymmetry $\epsilon$ is generated by  the complex phases in
 the Yukawa couplings $h_{ij}$.

During all these stages, the inflaton field continues to oscillate around 
the minimum of its potential and will eventually decay into SM degreees
of freeedom giving rise to the reheating stage. 
Before reheating, the universe is matter dominated because of the inflaton oscillations and the scale factor increases as $a \sim H^{-2/3}$. The lepton asymmetry $n_L \sim \epsilon n_{N_1}$ produced during the first oscillation at $H_{\rm osc}\sim {\widetilde m}/3$ is diluted at the time of reheating by the factor $a_{\rm osc}^3/a_{\rm RH}^3=H_{\rm RH}^2/H_{\rm osc}^2$. Expressing $n_{N_1}$ through \eq{nhu}, we find that the baryon asymmetry $Y_B=(8/23)(n_L/s)(H_{\rm RH}^2/H_{\rm osc}^2)$ becomes
\begin{equation}
\label{estimate}
Y_B\sim \frac{9~\epsilon ~h_t^{3/2}\,T_{\rm RH}\, |\phi_0|^{3/2}}{92\pi^3\widetilde{m}^{1/2}M_p^2}
= 10^{-6}\,\epsilon\,
\left( \frac{T_{\rm RH}}{10^7\,{\rm GeV}}\right)
\left(\frac{|\phi_0|}{M_p}\right)^{3/2} \left( \frac{100 \,{\rm GeV}}{\widetilde{m}}\right)^{1/2}.
\end{equation}
Notice that in our estimate we have not inserted any wash-out factor. Indeed,
as soon as the RH neutrinos decay, their energy density $\rho_{N_1}=M_1 n_{N_1}$ 
 gets promptly converted
into a ``thermal'' bath with an effective temperature $\widetilde{T}\sim 
(30\rho_{N_1}/g_*\pi^2)^{1/4}$ where $g_*$ is the corresponding 
number of relativistic degrees of freedom. We estimate that $\widetilde{T}$ is smaller than $M_1$ when
\beq
\label{ff}
M_1 > 10^9~{\rm GeV} \left(\frac{|\phi_0|}{M_p}\right)^{1/2}\left( \frac{\widetilde{m}}{100 \,{\rm GeV}}\right)^{1/2}.
\eeq
 As much heavier 
RH neutrinos are generated through 
the preheating stage,  we may safely conclude that $\Delta L=1$
inverse decays are not taking place. Similarly, one can show that
the  $\Delta L=2$ processes are out-of-equilibrium. 
Finally, flavour effects \cite{f1} play no role in determining the 
final baryon asymmetry as $\Delta L=1$ inverse decays are
out-of-equilibrium.  The maximum CP asymmetry
parameter for normal hierarchical light neutrinos, in the supersymmetric case, is given by $\epsilon=3 M_1 m_3/(4\pi \langle H_U\rangle^2)$, where $m_3=(\Delta m^2_{\rm 
atm})^{1/2}$ is
the largest light neutrino mass.
From Eq. (\ref{estimate}), we therefore estimate that 
enough baryon asymmetry is generated if 
\begin{equation}
\label{final1}
M_1\simgt 2 \times10^{11}\, {\rm GeV} \left( \frac{10^7\,{\rm GeV}}{T_{\rm RH}}\right)
\left(\frac{M_p}{|\phi_0|}\right)^{3/2} \left( \frac{\widetilde{m}}{100 \,{\rm GeV}}\right)^{1/2}.
\end{equation}
This limit, together with the result in \eq{mmax}, implies that a successful baryogenesis can occur only if $\phi_0\gsim 0.2\, M_p\, (10^7\,{\rm GeV}/T_{\rm RH})^{1/2}$. The condensate of the flat direction has to start its oscillation from field values close to the reduced Planck mass. Notice that this limit on $\phi_0$ is independent of $h_t$. However, the presence of the top Yukawa coupling is necessary to guarantee that the flat direction decays abundantly into $H_U$.

We conclude with some remarks. First, gravitinos are produced also during the 
instant preheating phase by scatterings of the quanta generated at the first
oscillation of the condensate. It is easy to estimate that their abundance is $n_{3/2}/s\simeq
10^{-4} (T_{\rm RH}/M_p)(\phi_0/M_p)^3$ and therefore it is never larger than the gravitino abundance 
produced at rehating by thermal scatterings, given by $n_{3/2}/s\simeq
2\times 10^{-12} (T_{\rm RH}/10^{10}~{\rm GeV})$. 
Secondly, from eq. (\ref{final1}) we infer that large values of the lightest RH neutrino mass $M_1$
are needed for the generation of a sufficiently large baryon asymmetry. However, we would like to
point out that our mechanism can work also in models with smaller values of $M_1$, since
the baryon asymmetry could be generated by the decays of the heavier 
RH neutrinos. Indeed, the up-Higgs may decay into the RH neutrinos $N_{2}$ (or $N_3$) instead
into the lightest RH neutrino $N_1$ if the condensate reaches the  value $\phi=\phi_{N_2}\equiv M_{2}/h_t$ before
the up-Higgs decays into $N_1$'s plus leptons. The time needed for the condensate
to reach the value $M_2/h_t$ is $\Delta t_{N_2}\sim \phi_{N_2}/\dot{\phi}\sim (M_2/h_t\widetilde{m} \phi_0)$ and 
is smaller  than the decay time of the up-Higgs into   $N_1$'s plus leptons 
if $\phi\lsim (8\pi\widetilde{m}\phi_0/
\sum_j|h_{1j}|^2 M_2)$. Imposing that this critical value is larger 
 than $M_2/h_t$, we find
that the up-Higgs will promptly decay into $N_2$'s rather than $N_1$'s if 
\begin{equation}
M_2\lsim 
\left(\frac{8\pi h_t\widetilde{m}\phi_0}{\sum_j|h_{1j}|^2}\right)^{1/2}.
\end{equation}
This condition can be satisfied if the Yukawas $h_{ij}$ 
are hierarchical and $|h_{1j}|\ll 1$. If this is the case, 
one should replace $M_1$ with $M_2$ (or $M_3$) in eqs. (\ref{ff}) and 
(\ref{final1}). 

In conclusion,  the observed baryon asymmetry can be explained within the 
supersymmetric leptogenesis scenarios  for low reheating temperatures 
and a RH hierarchical mass spectrum, thus avoiding the gravitino bound, if
two conditions are met: the initial value of the flat direction is
close to Planckian values, and the
phase-dependent terms in the flat direction potential are either vanishing
or sufficiently small for the 
particle
production to happen efficiently.

\centerline{\bf Ackowledgements}
\noindent
This research was supported in part by the European Community's Research
Training Networks under contracts MRTN-CT-2004-503369, MRTN-CT-2006-035505,  MRTN-CT-2006-035863  and MEST-CT-2005-020238-EUROTHEPHY (Marie Curie Early Stage Training Fellowship).

\end{document}